\documentclass[%
reprint,
superscriptaddress,
showpacs,preprintnumbers,1st_Draft.tex
bibnotes,
amsmath,amssymb,
aps,
pra,
floatfix
]{revtex4-1}

\usepackage{graphicx}
\usepackage{dcolumn}
\usepackage{bm}
\usepackage{braket}
\usepackage[utf8]{inputenc}
\usepackage[percent]{overpic}
\usepackage{xcolor}
\usepackage{amsmath}
\usepackage{subfigure}
\usepackage{dsfont}
\usepackage{bbm}

\begin{document}


\title{The role of quantum work statistics in many-body physics}

\author{J. Goold}
\affiliation{School of Physics, Trinity College Dublin, Dublin 2, Ireland}
\author{F. Plastina}
\affiliation{Dip. Fisica, Universit\`{a} della Calabria, I-87036 Arcavacata di Rende (CS), Italy}
\affiliation{INFN - Gruppo collegato di Cosenza, Cosenza, Italy}
\author{A. Gambassi}
\affiliation{SISSA --- International School for Advanced Studies, via Bonomea 265, I-34136 Trieste, Italy}
\affiliation{INFN, sezione di Trieste, via Bonomea 256, I-34136 Trieste, Italy}
\author{A. Silva}
\affiliation{SISSA --- International School for Advanced Studies, via Bonomea 265, I-34136 Trieste, Italy}

\date{\today}

\newcommand{\bracket}[2]{\langle{#1}|{#2}\rangle}
\newcommand{\ketbra}[2]{|{#1}\rangle\langle{#2}|}
\newcommand{\proj}[1]{|{#1}\rangle\langle{#1}|}
\newcommand{\comment}[1]{*\textit{[{#1}]}*}
\newcommand{\trace}{{\rm Tr}}
\newcommand{\beq}{\begin{equation}}
\newcommand{\eeq}{\end{equation}}
\newcommand{\bea}{\begin{eqnarray}}
\newcommand{\eea}{\end{eqnarray}}
\newcommand{\grad}{\vec\nabla}

	\newcommand{\tr}[1]{\textrm{Tr} \left[ {#1} \right]} 

	\newcommand{\e}[1]{e^{ {#1}}} 

	\newcommand{\s}[2]{\sigma^{#1}_{#2}} 

	\newcommand{\integral}[3]{{\int^{#2}_{#3} \mathrm{d}{#1} \;}} 

	\newcommand{\be}{\begin{equation}}
	\newcommand{\ee}{\end{equation}}


\newcommand{\abs}[1]{\left\vert#1\right\vert}
\newcommand{\Tr}[1]{\text{Tr}\left\{#1\right\}}
\def\I {{\rm 1} \hspace{-1.1mm} {\rm I} \hspace{0.5mm}}
\def\Z {{\mathds Z}}
\newcommand{\rosso}[1]{\color[rgb]{0.6,0,0} #1}
\newcommand{\ag}[1]{\textcolor{blue}{\bf #1}}

\begin{abstract}
In this contribution, we aim to illustrate how quantum work statistics can be used as a tool in order to gain insight on the universal features of non-equilibrium many-body systems. Focusing on the two point measurement approach to work, we first outline the formalism and show how the related irreversible entropy production may be defined for a unitary process. We then explore the physics of sudden quenches from the point of view of work statistics and show how the characteristic function of work can be expressed as the partition function of a corresponding classical statistical physics problem in a film geometry. Connections to the concept of fidelity susceptibility are explored along with the corresponding universal critical scaling. We also review how large deviation theory applied to quantum work statistics gives further insight to universal properties. The quantum-to-classical mapping turns out to have close connections with the historical problem of orthogonality catastrophe: we therefore discuss how this relationship may be exploited in order to experimentally extract quantum work statistics in many-body systems.
\end{abstract}

\pacs{}
\maketitle


\section{Introduction}
Over the past decade or so, there has been a surge of interest in the non-equilibrium dynamics of closed quantum systems following a switch of a Hamiltonian parameter. This is primarily due to a series of spectacular experiments in ultra-cold atoms, whereby the high degree of isolation permits the study of coherent dynamics over timescales typically inaccessible in conventional condensed matter physics~\cite{bloch2008,polkovnikov2011,cazalilla2011}. These experiments have raised new fundamental questions in the realm of non-equilibrium statistical mechanics, but have also revived a number of important theoretical issues such as the relationship between thermalisation and integrability~\cite{polkovnikov2011,eisert2015,borgonovi2016,gogolin2016,alessio2016} and the universality of defect generation following evolution across a critical point~\cite{dziarmaga2010}.

Over the same period of time, there has also been a great deal of activity in the statistical mechanics community surrounding the development of stochastic thermodynamics~\cite{sekimoto2010} and the study of non-equilibrium fluctuation relations in both the classical~\cite{jarzynski2011,seifert2012} and quantum domain~\cite{esposito2009,campisi2011,hanggi2015}. Loosely speaking, the name of the game is to study the thermodynamics of both classical and quantum systems beyond the linear response and to describe and understand the usual thermodynamic quantities such as work, heat and entropy as stochastic variables described by probability distributions. The fluctuation relations, then, encode the full non-linear response of a system to a time-dependent change of a Hamiltonian parameter. One particular feature is that the formalism permits the definition of irreversible entropy production of a unitary evolving system following a thermodynamic transformation; and, as such, it allows us to understand the emergence of thermodynamic behaviour in systems where the microscopic laws are inherently reversible~\cite{dorner2012}. These ideas have been cross-fertilized by the emergence of another community studying what has become known as quantum thermodynamics~\cite{goold2016} aiming at understanding the relationship between quantum mechanics and thermodynamics from first principles.

Given the current experimental interest in the non-equilibrium dynamics of quantum many-body systems, and the recent developments in statistical mechanics, along with the emergence of a flourishing community in quantum thermodynamics, it is natural to study the  dynamics of quantum many-body systems in this far-from-equilibrium thermodynamical formulation. In fact, this endeavour has been initiated a decade ago by Silva, who focused on explicit calculations of work statistics in a quantum critical many-body system~\cite{silva2008}; the universal features that can be uncovered in this way were further elucidated in a series of subsequent works~\cite{gambassi2011,gambassi2012, shchadilova2014,smacchia2012, smacchia2013,Kolodrubetz2013,palmai2015}. In fact, throughout this decade, there has been quite a remarkable amount of activity uncovering the features of work statistics in a range of physical models including spin chains~\cite{dorosz2008,dorner2012,mascarenhas2014,fusco2014,zhong2015work,apollaro2015,zhong2015,hoang2015, sharma2015, bayocbob2015, mazza2015,bayat2016nonequilibrium}, Fermionic systems~\cite{goold2011,Heyl2012b,Heyl2012a,knap2012,plastina2013decoherence,sindona2013orthogonality,campbell2014quenching,Schiro2014,Sindona2014,schiro2014transient}, Bosonic systems and Luttinger liquids~\cite{Roux2009,Dora2012,Sotiriadis2013,Dora2013, Bacsi2013,Dechiara2015,johnson2016thermometry,lena2016work,villa2018cavity}, periodically driven quantum systems~\cite{Bunin2011,Russomanno2015,Dutta2015,Lorenzo2017} among many others~\cite{Paraan2009,Wisniacki2013,Palmai2014,Gong2014,Deffner2015,liu2016quantum, eth,Jordi2017,cosco2017nonequilibrium,rotondo2018singularities}. Work statistics have also proved to be useful in the analysis of dynamical quantum criticality~\cite{Heyl2013,heyl2018} and more recently to shed light on the phenomenon of information scrambling~\cite{campisi2017,chenu2017}. The purpose of this brief review, together is to motivate and pontificate on the view that such non-equilibrium manipulations of quantum many-body systems can be seen, primordially, as thermodynamic transformations. In particular, we would like to focus our efforts on singling out what we consider to be the advantages of this line of reasoning and to highlight some interesting features of work statistics in many-body physics, which may not be apparent or appreciated across different communities. For the purposes of this contribution, we will primarily focus on the paradigm of sudden quenches.

We shall begin with an overview of work statistics and associated quantities such as the irreversible entropy production. We then move on to the issue of sudden quenches and show how the characteristic function is related to the partition function of  a higher-dimensional statistical model. From here, we show how it is possible to understand universal features of quench problems through connections with the well-known concepts of fidelity, fidelity susceptibility and large deviation theory~\cite{touchette2009}. Furthermore, we highlight the connection with the historically important problem of Anderson orthogonality catastrophe~\cite{anderson1967infrared} and the closely related Fermi edge singularity~\cite{mahan1967,nozieres1969} and explain how ongoing experiments in ultra-cold atoms are, in fact, linked to this problem and in principle can and should be used as a platform in order to extract work statistics in the many-body domain.

\section{Quantum Work Statistics and Thermodynamics}

Consider a quantum system described by a Hamiltonian $H(\lambda)$ that depends on an external work parameter $\lambda$, i.e. an externally controlled parameter whose value determines the equilibrium configuration of the system.
The system is prepared at time $t\le 0$ by allowing it to equilibrate with a heat reservoir at inverse temperature $\beta$ for a fixed value of the time-dependent work parameter $\lambda(t\leq0)=\lambda_0$. The initial state of the system is, thus, the Gibbs state
$\rho_\textrm{G}(\lambda_0)$, where
\begin{equation}
\rho_\textrm{G}(\lambda):=\frac{\e{-\beta H(\lambda)}}{\mathcal{Z}(\lambda)},
\label{eq:tmanifold}
\end{equation}
and the partition function $\mathcal{Z}(\lambda):=\tr{\e{-\beta H(\lambda)}}$. At time $t=0$ the system-reservoir coupling is removed and a protocol is performed on the system taking the work parameter $\lambda(t)$ from its initial value $\lambda_0$ to a final value $\lambda_\tau$ at a later time $t=\tau$.
The initial and final Hamiltonians connected by the protocol $\lambda_0\to\lambda_\tau$ have the spectral decompositions $H(\lambda_0)=\sum_n \epsilon_n(\lambda_0) \ket{\epsilon_n}\bra{\epsilon_n}$ and $H(\lambda_\tau)=\sum_m \epsilon'_m(\lambda_\tau) \ket{\epsilon'_m}\bra{\epsilon'_m}$, respectively,
where $\ket{\epsilon_n}$ ($\ket{\epsilon'_m}$) is the $n$-th ($m$-th) eigenstate of the initial (final) Hamiltonian with eigenvalue $\epsilon_n$ ($\epsilon'_m$). Work in the quantum domain results from a process and is not an observable in the sense that one cannot ascribe a Hermitian operator to it~\cite{talkner2007}.

The definition of the work done on the system as a consequence of the protocol, $W$, requires two projective measurements; the first projects onto the eigenbasis of the initial Hamiltonian
$H(\lambda_{0})$ at $t=0$, with the system in thermal equilibrium and renders a certain value $\epsilon_n$ with probability $p_n^0 = e^{-\beta \epsilon_n}/\mathcal{Z}(\lambda_0)$.
The system, then, evolves under the unitary dynamics $U(\tau; 0)$ generated by the protocol $\lambda_0\to\lambda_\tau$ before the second measurement
projects onto the eigenbasis $\{|\epsilon'_m\rangle \}$ of the final Hamiltonian $H(\lambda_\tau)$ and yields the values $\{\epsilon'_m\}$ with probability $p_{m|n}^\tau = |\langle \epsilon'_m|U(\tau,0)|\epsilon_n\rangle|^2$.
The probability of obtaining
$\epsilon_n$ for the first measurement outcome followed by $\epsilon'_m$ for the second measurement is then $p_n^0p_{m|n}^\tau$ 
and, accordingly, the work distribution is given by
\begin{equation}
P(W)=\sum_{n,m\ge 0} p^0_n\;  p^\tau_{m \vert n} \delta\left(W-(\epsilon_m'-\epsilon_n)\right).
\label{eq:qworkdist}
\end{equation}
Equation~\eqref{eq:qworkdist} therefore encodes the
fluctuations in the work that arise from both the thermal statistics
$p_n^0$ and the quantum measurement statistics $p^\tau_{m \vert n}$ over many identical realizations of the protocol. The first moment $\langle W\rangle$ of the distribution is the average work done and can be easily shown to be $\langle W\rangle=\tr{H(\lambda_{\tau})\rho_\tau}-\tr{H(\lambda_{0})\rho_{G}(\lambda_{0})}$ where $\rho_\tau=U(\tau,0)\rho_{G}(\lambda_{0})U^{\dagger}(\tau,0)$, i.e., nothing more than the energy change along the driven unitary process.

Now compare this transformation $\lambda_0 \to \lambda_\tau$ with an ideal quasi-static isothermal one, which, unlike an adiabatic transformation, is not unitary
in general, and
would bring the system through a path within the manifold of
equilibrium states described by Eq.~\eqref{eq:tmanifold}.
The work performed in the isothermal process is given by the free
energy change $\Delta F$. When this is subtracted from the actual work done
$\langle W\rangle$ one obtains back the so called 
irreversible work, which, when multiplied by the initial inverse temperature, defines the average irreversible entropy change
\begin{equation}
\langle S_{irr} \rangle = \beta\langle W_{irr} \rangle :=
\beta(\langle W \rangle - \Delta F) =   D(\rho_\tau || \rho_G(\lambda_{\tau}))
\label{sirr}
\end{equation}
The energetic deviation $\langle W_{irr} \rangle$ is often also
called dissipated work rather than irreversible. The reason it is
given the name irreversible is an assumption that
somewhere in the background there is a canonical thermal bath
where, after the driving, the system re-relaxes to a thermal state
at the same initial temperature.

The last equality in Eq. (\ref{sirr}) expresses the irreversible
entropy as the quantum relative entropy between the actual final
state $\rho_\tau$ and the final reference state $\rho_G (\lambda_{\tau})$,
$D(\rho_\tau || \rho_G(\lambda_{\tau}))= - S(\rho_\tau) - \Tr{\rho_\tau \ln
\rho_G(\lambda_{\tau}}$,  with $S(\rho)$ being the Von Neumann entropy
$S(\rho)=-\Tr{\rho\ln \rho}$. $D$ is the quantum analogue of the
Kullback-Leibler divergence and a very stringent measure of the
distinguishability of two quantum states via a
result known as quantum Stein's lemma. While not itself a metric,
it still upper bounds the trace distance via Pinsker's inequality,
\begin{equation}
\langle S_{irr} \rangle = D[\rho_\tau|| \rho_G(\lambda_{0})] \geq \|\rho_{\tau} -  \rho_G(\lambda_{0}) \|_1^2/2
\end{equation}
which captures the optimal distinguishability of quantum states with a single measurement. It can also be seen as a type of generalized second law for unitary processes~\cite{deffner2010}.
A non zero $\langle S_{irr} \rangle$, thus, signals the fact that the system has been brought out-of-equilibrium though a thermodynamically irreversible process, and it
also gives a quantification of how far from equilibrium it has gone, as it marks the difference between the actual final state $\rho_{\tau}$ and the equilibrium state $\rho_G(\lambda_{\tau})$. It is also directly connected to the fact that the work has a stochastic nature. Indeed, the existence of work fluctuations implies that the cumulants $C_n$ of the work distribution $P(W)$ are in general non-zero. In particular, the fact that $C_n\ne0$ with $n\geq 2$  means that $W$ has not a well defined value, as it is the case, instead, in the macroscopic thermodynamic context. It is possible to show that the irreversible entropy is, in fact, related to these cumulants by
\begin{equation}
\langle S_{irr} \rangle = \sum_{n=2} (-1)^n \frac{\beta^n}{n !} C_n \, ,
\end{equation}
which, in the linear response regime, under a gaussian approximation, reduces to $\langle S_{irr} \rangle = \beta^2 \sigma^2/2$, where $\sigma^2=C_2$ is the variance of the work distribution function.

Being cast in the form of a relative entropy, the strict positivity of $\langle S_{irr}\rangle$ is guaranteed via Klein's inequality. In fact, one could also work directly with the fluctuation theorems to demonstrate this positivity - for example Jarzynski's equality simply states that
$\langle e^{-S_{irr}}\rangle=1$ and then by application of Jensen's inequality one gets $\langle S_{irr}\rangle\ge 0$. Quantum fluctuation theorems and their important physical consequences are covered in many excellent and comprehensive overviews~\cite{esposito2009,campisi2011,hanggi2015} and we direct the reader towards them for more in depth analysis. For more information on work statistics we recommend the original paper detailing the non-observable nature of work~\cite{talkner2007} and an excellent overview paper on many aspects of the quantum work distributions~\cite{talkner2016aspects}.

\section{Sudden quench from the ground state}

\subsection{Quantum-Classical Correspondence and Universality}
\label{sec:silva}

One particular type of protocol that is very popular in ultra cold atomic experiments is the
so-called sudden quench, in which the change in the work parameter $\lambda$ is performed on a vanishingly small time scale. A very appealing thermodynamics description of such processes is given in terms of the characteristic function of work.

Consider, first, the characteristic function of the work distribution as the Fourier transform of Eq.~\eqref{eq:qworkdist} for a general time-dependent driving process,
\begin{align}
g(u,\tau) &:= \integral{W}{}{}\e{iuW}P(W),
\nonumber \\
&=\tr{U^\dag(\tau,0)\e{iuH(\lambda_\tau)}U(\tau,0)\e{-iuH(\lambda_0)}\rho_\textrm{G}(\lambda_0)}.
\label{eq:loschmidt}
\end{align}
In the case of a sudden quench, $\lambda_{0}\rightarrow\lambda_{f}$, the expression simplifies due to $U(\tau=0,0)=\mathds{1}$ such that $g(u,\tau=0)=g(u)=\tr{\e{iuH(\lambda_f)}\e{-iuH(\lambda_0)}\rho_\textrm{G}(\lambda_0)}$. This expression is very similar, in particular in the case of a pure initial state, to the core hole Green's function, quantity typically studied in the context of X-ray Fermi Edge singularities in condensed matter physics (see Sec. IV).

We now assume that the system is prepared in the ground state $|\epsilon_{0}\rangle$ of $H(\lambda_{0})$, i.e., $\rho_\textrm{G}(\lambda_0) = |\epsilon_0\rangle\langle\epsilon_0|$  and a sudden quench $\lambda_0 \to \lambda_f$ is performed so that the characteristic function now takes the form
\begin{align}
g(u) &= e^{-i\epsilon_0 u} \langle \epsilon_0| e^{iH(\lambda_f)u}|\epsilon_0 \rangle\\
&=\sum_{m\ge 0}e^{i(\epsilon^{\prime}_{m}-\epsilon_{0})u}|\langle \epsilon^{\prime}_{m}| \epsilon_{0}\rangle|^{2}.
\label{eq:lem}
\end{align}
If we interpret the conjugate variable $u$ as a time scale, we see that this expression represents, up to a phase,
a vacuum persistence amplitude, i.e., the probability amplitude to remain in the initial ground state at time $t=u$. As such, it is obviously
related to the survival probability $L(t)=|g^{*}(t)|^2$ which is often studied in quantum
chaos~\cite{peres1984stability}.
Most importantly, with the characteristic function written as a matrix element of the evolution operator $\exp[iH(\lambda_f)u]$ we can map upon analytic continuation
$u \rightarrow iR$ the sudden quantum quench problem to the problem of a classical system in a film geometry of thickness $R$. This was first pointed out by Silva in Ref.~\cite{silva2008}, but here we will outline the approach fleshed out in a later work~\cite{gambassi2011}.

Let us define the difference $\Delta\epsilon_{0}=\epsilon^{\prime}_0-\epsilon_0$ between the ground state energies of the pre- and post-quench Hamiltonians and perform the analytic continuation to the imaginary axis mentioned above, obtaining
\begin{align}
g(R)&=e^{-\Delta\epsilon_{0} R}\times \mathcal{Z}(R),\nonumber \\
\mathcal{Z}(R)&=\langle \epsilon_{0}|e^{-[H(\lambda_{f})-\epsilon'_{0}]R}|\epsilon_{0}\rangle.
\label{eq:partition}
\end{align}
%
%
%
For a $d$-dimensional quantum system possessing a $d+1$ dimensional classical correspondent, $\mathcal{Z}(R)$ can be seen as the partition function of the latter on a film of thickness $R$, with two boundary states $|\epsilon_{0}\rangle$ and a transverse ``area'' $L^d$ set by the extension of quantum system, assumed to be characterised by the large length $L$.
With the partition function form now in place, we can appeal to traditional statistical mechanics and take the logarithm of the expression in order to get a free energy $F(R)=-\ln(g(R))$ of the film and its corresponding density per transverse area, i.e., $f(R) = F(R)/L^d$. For large $R$ it is possible to separate this out into three contributions based on their dependence on decreasing powers of $R$, i.e.,
\begin{equation}
f(R)= R\times f_{b}+2f_{s}+f_C(R),
\label{eq:free}
\end{equation}
where $f_{b}=\Delta\epsilon_{0}/(R\times L^{d})$ is the bulk free energy density of the classical system and $f_{s}$ is the surface free energy per unit area associated with the two identical boundaries of the film. The remaining contribution $f_{C}(R)$ represents an effective finite-size interaction per unit area between the two confining surfaces which generically decays to zero at large separation $R$. In particular, one can write
\begin{equation}
\mathcal{Z}(R)=e^{-L^{d}[2f_{s}+f_C(R)]}.
\end{equation}

Let us now translate the information contained in the three components of the free energy density $f(R)$ into information about the statistics of the work. As stated above, the bulk free energy determined by $f_b$ is just related to a global phase in front of the vacuum persistence amplitude. Since $f_C(R) \rightarrow 0$ as $R\rightarrow +\infty$, the surface component $f_s$ of the free energy is, instead,  connected to the limit for large $R$ of the matrix element $\langle\epsilon_{0}|e^{-i[H(\lambda_f)-\epsilon^{\prime}_0]R} |\epsilon_{0} \rangle$ which defines ${\mathcal Z}(R)$ in Eq.~\eqref{eq:partition}. It is then easy to see that
\begin{equation}
e^{-2L^d\;f_s}=|\langle\epsilon_{0}|\epsilon^{\prime}_{0}\rangle|^2={\cal F}^2,
\end{equation}
where the quantity ${\cal F}$ introduced on the right-hand side is the so-called fidelity between the ground states of the post- and pre-quench Hamiltonians,
a quantity which is intensively studied in quantum information and many-body physics. In particular, it is a useful tool for analysing quantum critical systems~\cite{venuti2007quantum,zanardi2007information,gu2010fidelity}.
In the context of the work distribution for the sudden quench protocol at zero temperature, the fidelity ${\cal F}$ is the probability to measure the adiabatic work $W=\Delta\epsilon_0$ in the two-time measurement scheme. If the post-quench Hamiltonian $H(\lambda_f)$ of the quantum many-body system in question has a critical point at $\lambda_f=\lambda_c$, then the generalised susceptibilities~\cite{venuti2007quantum} $\chi_{n}(\lambda_{0},\lambda_{f})=-L^{-d}\partial^{n}_{\lambda_f}\ln{\mathcal{F}(\lambda_{0},\lambda_{f})}$,
%
%
can develop
a non-analytic behaviour~\cite{venuti2007quantum}. In particular, it is known that the fidelity susceptibility $\chi_{2}$ scales as
\begin{equation}
\label{eq:fidelitys}
\chi_{2}\propto|\lambda_f-\lambda_{c}|^{\nu d-2} \, ,
\end{equation}
where $\nu$ is the correlation length critical exponent.

The generalised susceptibilities $\chi_n$ have a straightforward interpretation in terms of the analogy with boundary statistical mechanics, which is particularly suggestive when the quenched parameter $\lambda$ of $H(\lambda)$ can be interpreted as a "temperature" in the $d+1$ classical correspondent of the quantum model. This is the case, for example, of the transverse field in the quantum Ising chain~\cite{sachdev}.
Since in general the generalized susceptibilities can be written as derivatives of the surface free energy,
\begin{equation}
\label{eq:fidelitys2}
\chi_{n}(\lambda_0,\lambda_f)=\partial^{n}_{\lambda_{f}}f_{s}(\lambda_0,\lambda_{f}).
\end{equation}
%
%
it is evident that in when $\lambda$ is a temperature-like variable they have a clear physical interpretation, $\chi_{1}$ being the excess internal energy and $\chi_2$ the excess specific heat of the corresponding $d+1$ classical confined system. The excess specific heat is well-known to scale as $\chi_2\propto |\lambda_f -\lambda_c|^{-\alpha_{s}}$ close to the critical point where the exponent $\alpha_s$ is related to the bulk critical exponents of the correlation length $\nu$ and of the specific heat $\alpha$ by $\alpha_{s}=\alpha+\nu$ of the classical $d+1$ dimensional system and satisfy the hyperscaling relation $\alpha+\nu(d+1)=2$ \cite{D-86,D-861}.

Relating quantum quenches to the statistical physics of classical confined systems provides useful information not only on the singularities of the generalised susceptibilities, but also on the emergent universal features of the probability distribution function of the work when, as above, the post-quench Hamiltonian is close to a critical point. Before proceeding, we note that for a quench starting from the ground state $|\epsilon_0\rangle$ of the pre-quench Hamiltonian, the quantity $\Delta\epsilon_0$ represents the \emph{reversible} work one would do on the system by performing the change $\lambda_0\to\lambda_f$ adiabatically. This also sets the minimal possible value of the stochastic variable $W$ and, accordingly, the irreversible work $W_{irr} = W - \Delta\epsilon_0$ can take only positive values. This means that its probability distribution $P(W_{irr})$ displays a lower edge and, based on the correspondence highlighted above, we can finally demonstrate that
\begin{equation}
\label{genericpw}
\begin{split}
P(W_{irr})\simeq {\cal F}^2 \, [&\delta(W_{irr}) \\
&+{\cal C}\,(W_{irr}-q m)^{1-a}u_s(W_{irr}-q m) + \dots],
\end{split}
\end{equation}
where $u_s(\cdots)$ is the Heaviside unit step and the parameters appearing in this expression are the fidelity ${\cal F}$ (defined above), the mass $m$ of the lightest quasi-particle in the model, and three \emph{universal}  constants ${\cal C}$, $q$, and $a$.
%
%
%


This universality stems ultimately from the fact that the leading part of the finite-size free energy $f_C(R)$ of the $d+1$-dimensional classical systems in finite-size geometry \cite{FSS,FSS1} acquires a universal character whenever it --- and thus the post-quench Hamiltonian $H(\lambda_f)$ of the $d$-dimensional quantum system ---  is close to a bulk \emph{critical point}. In particular this so-called critical Casimir effect \cite{G-09} is characterised by a scaling behaviour
\begin{equation}
f_{C}(R) = R^{-d} \Theta_{\mathcal B}(\pm R/\xi_+),
\label{eq:CCF}
\end{equation}
where $+$ and $-$ refer to the disordered and ordered phases, respectively. $\xi_+ \propto |\lambda_f-\lambda_c|^{-\nu} $ is the exponential correlation length associated with the critical fluctuations of the relevant order parameter of the transition, which grows upon approaching it. Both $\xi_+$ and $R$ are assumed to be much larger than any microscopic length scale of the system such as, e.g., a possible lattice spacing.
The scaling function $\Theta_{\mathcal B}$ is characterised by a certain degree of \emph{universality} \cite{G-09,G-10}, as many other properties close to critical points, which become independent of the microscopic features of the system. In particular, $\Theta_{\mathcal B}$ depends only on the universality class of the classical critical point and, because of the presence of the boundaries set by the quantum state $|\epsilon_0\rangle$, it depends also on their \emph{surface universality class} \cite{D-86,D-861}  or, equivalently, on which of the few effective boundary states ${\mathcal B} \in  \{|\phi^*_i\rangle\}_i$, the initial/boundary state $|\epsilon_0\rangle$ flows to as the critical point is approached.  Accordingly, $\Theta_{\mathcal B}$ is also largely independent of the specific values assumed by $\lambda_0$ and $\lambda_f\simeq \lambda_c$.
In view of its numerous applications to the physics of soft matter, the critical Casimir effect has been extensively studied both theoretically and experimentally in the past few years (see, e.g., Ref.~\cite{G-09}) and many of its features are known, including the scaling function $\Theta_{\mathcal B}$ for a variety of bulk and surface universality classes.

While the large-$R$ decay of $f_C$ is $\propto R^{-d}$ at criticality $\xi_+ = \infty$, away from the critical point it is dictated by the asymptotic expansion of $\Theta_{\mathcal B}(x)$ for $|x|\gg 1$. Generically, it takes the form
\begin{equation}
\label{asympt}
\Theta_{\mathcal B}(x\rightarrow \pm \infty) = {\cal C}_\pm |x|^{a_\pm}\; e^{-q_\pm |x|}+ \dots,
\end{equation}
where ${\cal C}_\pm$, $a_\pm$ and $q_\pm>0$ are universal constants (dependent only on ${\mathcal B}$), which take different values within the ordered ($-$) and disorderd $(+)$ phases.
%
For the quantum (classical) Ising model in $d=1$ ($d=2$) one has $q_+=a_+=1$, while $q_-=2$ and $a_-=-1/2$, corresponding to two possible instances of quenches; i.e., within the same phase or across the quantum phase transition.
These constants directly enter Eq.~\eqref{genericpw}, while the mass $m$ of the lightest quasi-particle of the quantum model in the paramagnetic phase has to be identified with the inverse $\xi_+^{-1}$ of the correlation length of the classical system.
%
%
Performing the analytic continuation $R\mapsto -iu$ the behavior of $P(W)$ for $W$ close to threshold is determined by the asymptotic behavior
of the characteristic function $g(u)$ for large $u$. Therefore, performing a large $u$ expansion of $g(u)$ and 
taking its Fourier transform we readily obtain that $P(W)$ takes the form in Eq.~\eqref{genericpw}.



\subsection{Large Deviations}

Work is an extensive quantity in thermodynamics. Standard statistical mechanics textbook tells us that the mean of a generic extensive quantity such as the average work $\langle W\rangle$ done on a system will grow proportionally to its number $N$ of degrees of freedom, where $N=L^d$ for a system of typical size $L$ in $d$ spatial dimensions.
Accordingly, it is natural to define an associated \emph{intensive} quantity $w$ by dividing $W$  by $N$ and, assuming weak correlations among the $N$ degrees of freedom, the central limit theorem tells us that the distribution of $w$ will be generically Gaussian for large $N$, with fluctuations $\Delta w =\sqrt{ \langle (w-\bar w)^2\rangle}$ around the average value $\bar w \equiv \langle w \rangle$ suppressed as $1/\sqrt{N}$. This means that, generically, $w-\bar w \sim N^{-1/2}$ and that the distribution of $w$ concentrates around its average and most probable value $\bar w$.
However, rare as they may be, large deviations of $w$ with $w-\bar w \sim 1$ are well-known to be able to probe specific features of statistical systems~\cite{touchette2009}. As shown in
Ref.~\cite{gambassi2012}, the work statistics of a sudden quantum quench can be cast in the framework of large deviation theory and it might display some \emph{universal} features when the post-quench Hamiltonian is close to an equilibrium critical point.

Consider now as the stochastic variable the intensive  work $w=W/N$ as opposed to extensive one $W$. As shown in Ref.~\cite{gambassi2012}, the probability $P(w)$ that a large fluctuation will occur is expected to be exponentially small in the size $N$, i.e.,
\begin{equation}
P(w)\propto\exp[-NI(w)],
\label{eq:largedev}
\end{equation}
where $I(w)$ is the non-negative rate function which characterises the large deviations and which vanishes for $w=\bar w$. The quadratic approximation of $I(w)$ around $w=\bar w$ describes the typical Gaussian fluctuations $w-\bar w \sim N^{-1/2}$ expected on the basis of the central limit theorem.
Let us consider the case of a quench from the ground state $|\epsilon_0\rangle$ of the pre-quench Hamiltonian: as discussed in the previous section, the
%
irreversible work $W_{irr} = W - \Delta\epsilon_0$ can take only positive values. For convenience, we focus on the large deviations of the irreversible intensive work $W_{irr}/N$, which is denoted below, for simplicity, by $w$, with $P(w< 0)=0$ and therefore $I(w<0) = +\infty$.
In order to compute $I(w)$ appearing in Eq.~\eqref{eq:largedev} it is convenient first to focus on the moment generating function of $W_{irr}$, i.e., on $\langle e^{- R W_{irr}}\rangle$ which is related to $g(R)$ in Eq.~\eqref{eq:partition}. In particular, because of the shift in the definition of $W_{irr}$ compared to $W$, the contribution $N R \times f_b$ corresponding to the bulk free energy on the r.h.s.~of that equation is cancelled, and only the so-called \emph{excess free energy} $N f_{ex}(R)$, with density $f_{ex}(R)=f(R)-Rf_b$,  contributes to the moment generating function \cite{gambassi2012}:
\begin{equation}
\langle e^{- R W_{irr}}\rangle = \exp[-N f_{ex}(R)].
\label{eq:moment}
\end{equation}
Note that the generating function on the r.h.s.~is certainly defined for all possible non-negative values of $R \in {\mathbb R}^+$ and it is actually related to the excess free energy density of the corresponding classical system in a film geometry $L^d\times R$, as discussed in Sec.~\ref{sec:silva}, only in this case. Depending on the behaviour of $P(W_{irr})$ for large $W_{irr}$, however, the domain ${\mathcal D}$ within which the generating function is defined can also include negative values of $R$ and therefore the equality on Eq.~\eqref{eq:moment} is understood after an analytic continuation of $f_{ex}(R)$ on the r.h.s.~towards $R<0$.
With the generating function in the form \eqref{eq:moment} we can apply the formalism of large deviation theory that gives us the prescription to evaluate the rate function through the Legendre-Fenchel transformation of $f_{ex}(R)$,
\begin{equation}
I(w)= - \inf_{R \in {\mathcal D}}[R\,w-f_{ex}(R)],
\label{eq:LF}
\end{equation}
i.e., via a saddle-point evaluation of $P(w)$ in Eq.~\eqref{eq:largedev} as the inverse Laplace transform of Eq.~\eqref{eq:moment} for $N\to\infty$.
From the r.h.s.~of Eq.~\eqref{eq:moment} one sees that $f_{ex}(0)=0$, $f'_{ex}(0)=\bar w$, and that $f_{ex}(R)$ is a concave function of $R$, being the exponential function on the l.h.s.~a convex function. In addition, $f_{ex}$ approaches $2f_{s}$ as $R\rightarrow\infty$.
We can infer some properties of the rate function $I(w)$ on the basis of these qualitative features of $f_{ex}$. For $w<0$, for example, the infimum on the r.h.s.~of Eq.~\eqref{eq:LF} is attained for $R\to\infty$ and therefore $I(w<0) = +\infty$, as expected because $W_{irr} \ge 0$. Similarly, the behaviour of $I(w)$ close to the threshold $w\to 0^+$ is determined by that of $f_{ex}(R\to\infty)$ and, in particular, $I(0) = 2 f_s$. The way this value is approached depends on the finite-size contribution $f_{C}(R)$ in Eq.~\eqref{eq:free}.
As discussed in Sec.~\ref{sec:silva}, this contribution acquires a universal character whenever the corresponding $d+1$-dimensional classical system --- and therefore the post-quench Hamiltonian $H(\lambda_f)$ of the $d$-dimensional quantum system ---  is close to a bulk critical point. In this case, $f_C(R)$ takes the universal scaling form in Eq.~\eqref{eq:CCF},
characterised by the scaling form $\Theta_{\mathcal B}$, which is known in a variety of cases  (see, e.g., Ref.~\cite{G-09}).

Based on this knowledge of $\Theta_{\mathcal B}$, the rate function $I(w)$ can be readily calculated via Eq.~\eqref{eq:LF}, finding
\begin{equation}
I(w) = 2 f_s + \xi^{-d} \vartheta(w\xi^{d+1}),
\end{equation}
where $\vartheta(y)$ is the Legendre-Fenchel transform (see Eq.~\eqref{eq:LF}) of $x^{-d}\Theta_{\mathcal B}(x)$ and it is as universal as $\Theta_{\mathcal B}$.

Accordingly, not only the edge singularities of the extensive work $W$ discussed in Sec.~\ref{sec:silva} above are determined by universal features of the system and of the quench, but also the large deviations of the intensive variable associated with the irreversible work $W_{irr}$ display universal properties in their rate function $I(w)$ close to the threshold $w=0$. For larger values of $w$, the rate function $I(w)$ depends on the excess free energy $f_{ex}(R)$ of films with increasingly smaller thickness $R$, which even becomes negative for $w\ge \bar w$. In this case, the correspondence with the physics of classical systems in film geometry breaks down and universality is generically lost. However, concrete examples show that
interesting phenomena analogous to a Bose-Einstein condensation \cite{gambassi2012} as well as non-analyticities of the rate function $I(w)$ \cite{rotondo2018singularities} similar to non-equilibrium phase transitions may occur.

\subsection{Thermal Quenches}

The previous two sub-sections, strictly speaking, describe only sudden quenches starting from the ground-state, which are characterised by the fact that the work $W$ cannot be smaller than the value $\Delta \epsilon_0$. Accordingly, the probability distribution $P(W)$ of $W$ features an edge which acquires the universal features discussed above. Here we will demonstrate that interesting information can still be obtained from both the average work and the irreversible entropy production when an initial thermal state $\rho_{G}(\lambda_{0})$ is assumed and the edge is absent. The average work done $\langle W\rangle=\int P(W)W dW$ by a sudden quench $\lambda_0\to \lambda_f$ of a parameter $\lambda$ which couples linearly in $H(\lambda)$ and starting from a thermal state can be written in the following form~\cite{Sotiriadis2013,mascarenhas2014}
\begin{align}
  \langle W\rangle &=\frac{\Tr{e^{-\beta H(\lambda_{0})}[H(\lambda_f)-H(\lambda_0)] }}{\Tr{e^{-\beta H(\lambda_{0})}}} \\ \nonumber
&=(\lambda_{f}-\lambda_{0}) F'_{\beta}(\lambda_{0}),
\label{eq:firstd}
\end{align}
which follows from the fact that $H(\lambda_f) - H(\lambda_0) = (\lambda_f-\lambda_0)\partial{H(\lambda_{0})}/\partial{\lambda_{0}}$ and equals the derivative of the equilibrium free energy with respect to the parameter $\lambda_{0}$ times the quench amplitude. From this, we can see that, for sudden quenches, the expression for the irreversible work takes the interesting form $\langle W_{irr}\rangle=\langle W\rangle-\Delta F=(\lambda_{f}-\lambda_{0}) F'_{\beta}(\lambda_{0})- F_{\beta}(\lambda_{f})+F_{\beta}(\lambda_{0})$. Let us restrict ourselves to small quenches, such that  $\delta\lambda=\lambda_f-\lambda_0\ll 1$; in this case, the irreversible entropy production defined by Eq.~\eqref{sirr} can be seen as proportional to the second derivative of the equilibrium free energy,
\begin{equation}
 \langle S_{irr}\rangle=-(\delta\lambda)^{2}\beta F''_{\beta}(\lambda_{0})/2+\mathcal{O}(\beta(\delta \lambda)^{3}).
\end{equation}
Generically, possible non-analytic behaviour in $F''_{\beta}$ is characterised by the critical exponent $\alpha$ of the specific heat $C\propto-F_{\beta}''(\lambda_{0})\propto|\lambda_{0}-\lambda_{c}|^{-\alpha}$. At finite temperature, there is no quantum phase transitions; but, as was shown in \cite{dorner2012} for quenches of the Ising model, the irreversible entropy production starts to diverge as the temperature decreases, thus signalling the proximity to a quantum critical point. In a first order quantum phase transition, the average work becomes discontinuous. A comparison between first and second order phase transitions has been performed in \cite{mascarenhas2014}. We note that we have been explicitly considering sudden quench problems in the preceeding sections and from the many body perspective very little work has been done on more generic time dependent processes~\cite{smacchia2013} it would be interesting to extend studies in this direction. In addition there has been recently some interesting work in the direction of finite time charging of quantum batteries with connection to quantum correlations~\cite{alicki,binder1,barcelona,binder2,pisa}, given that these studies are finite time manipulations of interacting quantum systems, it would be interesting to explore connections with the prescription outlined here.

\section{Anderson Orthogonality Catastrophe and current experiments}

\subsection{Orthogonality Catastrophe}
In this section, we would like to discuss how the concept of quantum work statistics in sudden quenches is actually connected to the historically important problem of orthogonality catastrophe (OC) in condensed matter physics. This phenomenon was discovered by Phil Anderson in 1967~\cite{anderson67}. Anderson was studying the seemingly innocuous problem of a single scatterer in the presence of a non interacting Fermi gas. He considered the ground states of both $N$ Fermions in a spherical box of radius $R$  in the presence and absence of a single local scattering potential with only s-wave ($l=0$) contribution. In the presence of a scattering potential, $V$, the single particle states in the box acquire a phase shift $\delta(E)$. The fermions being assumed to be non interacting, the ground states are expressed as Slater determinants of the single particle eigenstates. Let the ground state of the unperturbed system be $\Psi_{i}(x_{1},x_{2},\dots x_{N})$ and the ground state of the perturbed system be $\Psi_{f}(x_{1},x_{2},\dots x_{N})$; Anderson proved that the overlap (fidelity) of these two states scales as
\begin{align}
F &=\int dx_{1}dx_{2}\dots dx_{N}\Psi^*_{f}(x_{1},x_{2},\dots x_{N})\Psi_{i}(x_{1},x_{2},\dots x_{N})\\ \nonumber
&=N^{-\alpha},
\label{eq:oc}
\end{align}
where $\alpha=\delta^{2}/\pi^2$. This implies that the ground states become orthogonal as the system size increases with a power-law that depends universally on the phase shift $\delta$. This phenomenon is known as the orthogonality catastrophe (OC). Innocuous and all as this result may first seem, it actually has  several deep implications for the physics of Fermi gases. For example, there is an immediate consequence for the situation where a local impurity is made to interact with the gas. Generically, the interaction between an impurity and a Fermi gas will lead to a dressing of the impurity by the excitations of the gas. In particular, when the mass of the impurity is finite, then one may talk about the formation of a well defined quasi-particle: the Fermi polaron. In condensed matter physics, this effect is quantified by the quasi-particle residue, which, mathematically speaking, is equivalent to the fidelity between perturbed and un-perturbed ground states. In the limit of a heavy particle, instead, the problem may be framed in the infinite mass approximation and the dressed particle looses its quasi particle description as the fidelity goes to zero due to the manifestation of the OC.

At this point one might ask: what is the connection between work statistics and the previously recalled formalism?  Well, going beyond ground-state physics, perhaps the most dramatic consequence of the OC is in non-equilibrium dynamics. Historically, this consequence was discovered, very soon after the original realisation by Anderson, by Nozieres and de~Dominicis~\cite{nozieres1969} who were considering the many electron response to the sudden switching on of a core hole in a metal. Physically, this occurs after an x-ray photon has created a deep hole, with the promotion (emission) of a core electron in a metal. Nozieres and de~Dominicis considered the core hole Green's function, which is defined as
\begin{equation}
\mathcal{G}(t)=-ie^{-\omega_{T}}u_s(t)\langle e^{itH_{i}}e^{-it(H_{i}+V})\rangle,
\end{equation}
where $\omega_{T}$ is the threshold frequency for the creation of a core hole in the valence band, $u_s(t)$ is the Heaviside step function and $H_{i}$ and $H_{i}+V$ are the perturbed and unperturbed Hamiltonians respectively.  We note that for the purposes of illustration we are assuming an initial thermal state but the calculation can be performed directly in the zero temperature limit~\cite{mahan2010}. The key quantity of this problem is the vacuum persistence amplitude,
\begin{equation}
\nu_{\beta}(t>0)=\left\langle e^{\frac{i}{\hbar }\hat{H}_{i}t}e^{-\frac{i}{%
\hbar }\left( \hat{H}_{i}+\hat{V}\right) t}\right\rangle.
\label{Eq:VacAmp}
\end{equation}%
This gives the response of the gas to the switching on of the localized impurity potential, and, as pointed out previously, this coincides with the complex conjugate of the characteristic function of work defined by Eq.~\eqref{eq:lem}, thus giving an immediate connection to the work statistics formalism. We will return to this shortly. In the interaction picture, the expression for the vacuum persistence amplitude (or if you like, the complex conjugate of the characteristic function of work) reads:
\begin{equation}
\nu_{\beta}(t)=\left\langle Te^{-\frac{i}{\hbar }\int_{i}^{t}dt^{\prime }%
\tilde{V}(t^{\prime })}\right\rangle \text{,\quad }\tilde{V}(t)=e^{\frac{i}{%
\hbar }\hat{H}_{0}t}\hat{V}e^{-\frac{i}{\hbar
}\hat{H}_{0}t}, \label{eq:Vbeta}
\end{equation}%
which, by virtue of the linked cluster theorem, reduces to an
exponential sum of connected Feynman diagrams containing an increasing number of interaction vertices:\\
\scalebox{0.95}{\includegraphics{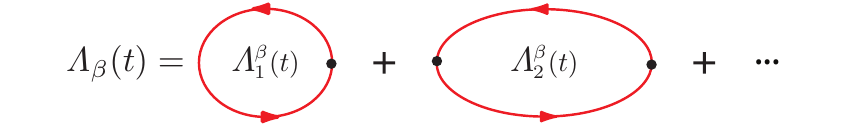}}\\
If the reader is interested in precise details of the calculation of the linked cluster expansion we recommend the text book by Mahan~\cite{mahan2010}. It turns outs that the first term in the sum is nothing but a first order shift to the Fermi gas energy, which brings only a phase factor to $\nu_{\beta}(t)$. The second term, instead, gives the dominant (singular) contribution, which is a direct manifestation of the OC. Taking the $\beta\rightarrow\infty$ limit, an explicit calculation gives
\begin{equation}
\Lambda_2^{\beta\rightarrow\infty}(t)=-g \, \ln(it/\tau_{0}+1), \label{historicMND}
\end{equation}
where $g$ is a rescaled impurity interaction strength parameter which is proportional to $\alpha$ in the Anderson overlap. The mathematical consequence of this result is a power-law decay of $\nu(t)$. However, the real physical implication becomes apparent if we consider the absorption spectrum, which is precisely what is measured in X-ray scattering experiments and is defined as
\begin{equation}
A(\omega)=2 Re \int^{+\infty}_{0}\!\!dt\, e^{i\omega_{T}t}\nu(t).
\end{equation}
The power-law decay of $\nu(t)$ leads to a power-law threshold singularity in the absorption spectrum. This iconic non-equilibrium effect is known as the Fermi edge singularity in condensed matter physics.
The physics is crucially encapsulated by the Anderson overlap and it is illuminating now to rather interpret the phenomenon from a thermodynamic perspective. Since $\nu_{\beta\rightarrow\infty}$ is nothing more than the complex conjugate of the characteristic function, then the absorption spectrum can be interpreted as the probability to do work on the system as they are mathematically equivalent. Writing the absorption spectrum in the Lehmann representation, indeed, we have
\begin{equation}
A(\omega)\propto \sum_{m}|\langle \epsilon'_{m}|\epsilon_{0}\rangle|^{2}\delta(W-\epsilon'_{m}+\epsilon_{0}).
\end{equation}
The thermodynamic implication of these considerations should be immediately obvious. Interpreting the creation of a core hole as thermodynamic work on a metal, and interpreting the absorption spectrum as the work distribution $P(W)$, we see that due to the OC there is no possibility that this process can be adiabatic. The probability to do adiabatic work $W=\epsilon'_{0}-\epsilon_{0}$ goes to zero as a power-law due to the OC between $|\epsilon_{0}\rangle$ and the ground state of the system. In \cite{sindona2015statistics} the authors have undertaken a very careful examination of the moments of the work statistics and found that it was in fact the third moment of the work distribution, which quantifies the skewness or asymmetry of the process, which scales with the universal exponent of the edge-singularity $g$. In summary, we would like to stress that, although the motivation is very different, from the operational perspective the mathematics of the Fermi-edge singularity problem at its core are identical to the formal framework needed to describe the quantum work statistics of a sudden {\it local} quench of a Fermi gas. This is, in fact, quite important and not just a mere curiosity, as it paves the way for a detailed experimental study of quantum work statistics in ultra-cold atom setups, where orthogonality catastrophe and Fermi-edge singularity physics are currently being probed. This connection is not fully appreciated in either the ultra-cold atom or the thermodynamics community. Indeed, it has been argued that trapped ultra-cold fermion atoms could constitute an almost ideal set-up to investigate this physics in a controlled fashion \cite{goold2011,knap2012,schmidt18}, and a detailed treatment of the Fermi edge singularity problem for a harmonically trapped gas has been recently performed \cite{sindona2013orthogonality,Sindona2014}, which generalises the result in Eq. (\ref{historicMND}).

\subsection{Experiments with Ultra-cold Fermions}

The experimental extraction of quantum work statistics involves the rather precarious setting of not only having to prepare a well defined initial state and perform controlled unitary operations, but also there is the apparent stringent necessity of performing two non destructive projective measurements on the eigen-basis of the system. This is in contrast to classical work statistics and has rendered the experimental acquisition of quantum work statistics elusive up until relatively recently. The first proposal was for a clever phonon shelving technique for direct extraction of work statistics in an ion trap~\cite{huber2008}. Actually as it turned out, two papers appearing at the same time proposed the use of Ramsey interferometry on an ancillary qubit in order to extract the characteristic function of work~\cite{Dorner2013,Mazzola2013} (see also \cite{Campisi2013}) avoiding the difficult projective measurements. This led to the first experimental extraction of work statistics and experimental verification of the fluctuation theorems in a quantum system in a liquid state NMR setup~\cite{Batalhao2014}. An experimental work confirming the first proposal for direct measurement in the energy domain appeared soon after in an ion trap setup~\cite{an2015experimental}. More recently, following the realization that a work distribution can in principle be reformulated as a positive operator valued measurement (POVM)~\cite{povm,Dechiara2015}, an experiment has been performed to reconstruct the distribution $P(W)$ with Rubidium atoms in an atom chip~\cite{cerisola2017using}.

More directly related to our discussion about the relations between work distribution and OC, two experiments have been performed with cold Li atoms, forming a Fermi sea to which an impurity K atom is coupled \cite{cetina15,cetina16}. After preparing the impurity in a superposition of its (two lowest) internal energy states, Feshbach resonance has been used to tune the coupling with the gas and to  switch it on only in the excited state. The theoretical blue print for this idea were first outlined in \cite{goold2011} and \cite{knap2012}. Using Ramsey interferometry of the impurity internal state and monitoring the decoherence dynamics, the vacuum persistence amplitude has been experimentally measured, both in amplitude and phase, together with the absorption spectrum, for both repulsive and attractive impurity-gas interactions.

Given the direct connection between work statistics and the OC and associated Fermi-edge singularity problem, we would like to strongly emphasise that these experimental platforms of dilute Fermi mixtures are ideal playgrounds for the controlled exploration of universal features in the quantum thermodynamics of many-body systems.

\begin{acknowledgements}
J.G. is supported by a SFI Royal Society University Research Fellowship. \end{acknowledgements}


\bibliography{review}
\end{document}